\documentclass{aa}
\usepackage[comma,authoryear]{natbib}
\usepackage{natbib}
\bibpunct{(}{)}{;}{a}{}{,}
\usepackage{epsfig}
\usepackage{graphics}
\usepackage{float}
\usepackage{amsmath}
\usepackage{multirow}
\usepackage{longtable}
\usepackage{rotate}
\usepackage{amssymb}
\def\sidehead#1{\noalign{\vskip 1.5ex}\multicolumn{4}{@{}l}{\em #1}\\
                \noalign{\vskip .5ex}}

\def\phn{\phantom{0}}  
\def\phs{\phantom{$-$}}    
\def\phn{\phantom{0}}
\def\nodata{\phs$\cdots$}
\def\kms{\ifmmode{\rm km\,s^{-1}}\else\hbox{$\rm km\,s^{-1}$}\fi}

\def\tablecomments#1{\par\smallskip\noindent Notes. #1}
\def\plotone#1{\centerline{\psfig{figure=#1,width=\hsize,clip=}}}
\let\arcdeg\degr
\let\simgt\ga
\let\simlt\la
\setlongtables

\let\internalcite\cite
\def\cite{\def\authoryear##1{##1}\internalcite}

\begin{document}

   \title{Time-resolved optical spectroscopy of the pulsating DA white dwarf
       HS 0507+0434B \thanks{The data presented herein were obtained at the 
       W.M. Keck Observatory, which is operated as a scientific partnership 
       among the California Institute of Technology, the University of California 
       and the National Aeronautics and Space Administration. The Observatory 
       was made possible by the generous financial support of the W.M. Keck 
       Foundation.}}
       
  \subtitle{New constraints on mode identification and pulsation properties}

   \author{R. Kotak\inst{1}
          \and{M. H. van Kerkwijk}\inst{2}
           \and{J. C. Clemens}\inst{3}\thanks{Alfred P. Sloan Research Fellow}}
   \offprints{R. Kotak}

   \institute{ Lund Observatory,
               Box 43, SE-22100 Lund, Sweden\\
               \email{rubina@astro.lu.se}
           \and
              Astronomical Institute, Utrecht University,
              P. O. Box 80000, 3508~TA Utrecht, The Netherlands\\
              \email{M.H.vanKerkwijk@astro.uu.nl}
           \and 
              Department of Physics and Astronomy, University of
              North Carolina, Chapel Hill, NC 27599-3255, USA\\
              \email{clemens@physics.unc.edu}
             }

   \date{Received ; accepted }

   \abstract{
We present a detailed analysis of time-resolved optical spectra 
of the ZZ Ceti white dwarf, HS 0507+0434B. 
Using the wavelength dependence of observed mode amplitudes, we
deduce the spherical degree, $\ell$, of the modes, most of which
have $\ell=1$. The presence of a large number of combination
frequencies (linear sums or differences of the real modes) enabled
us not only to test theoretical predictions but also to indirectly 
infer spherical and azimuthal degrees of real modes that had no 
observed splittings. In addition to the above, we measure line-of-sight 
velocities from our spectra. We find only marginal evidence for 
periodic modulation associated with the pulsation modes: at the frequency 
of the strongest mode in the lightcurve, we measure an amplitude of 
$2.6\pm1.0$\,kms$^{-1}$, which has a probability of 2\% of being due 
to chance; for the other modes, we find lower values. Our velocity amplitudes 
and upper limits are smaller by a factor of two compared to the amplitudes 
found in ZZ Psc. We find that this is consistent with expectations based 
on the position of HS 0507+0434B in the instability strip. 
Combining all the available information from data such as ours is a first 
step towards constraining atmospheric properties in a convectionally 
unstable environment from an observational perspective.
   }
   \authorrunning{Kotak}
   \titlerunning{The pulsating DA white dwarf HS 0507+0434B}
   \maketitle

\section{Introduction} 
The apparent simplicity of white dwarfs belies more complex and
ill-understood processes occurring in their interiors. 
Pulsations, however, offer the precious possibility of
probing the interiors of these stars thereby yielding not only 
fundamental stellar parameters but also important clues of 
their prior history which can, in principle, be reconstructed from
their present interior structure.

Along the white dwarf cooling track, there are three regions of instability: 
at $T_{\mathrm{eff}}\sim$ 110\,kK, $T_{\mathrm{eff}}\sim$ 24\,kK,
and $T_{\mathrm{eff}}\sim$ 12\,kK, populated by the GW Vir, V 777 Her 
(DBV), and ZZ Ceti (DAV) types, displaying strong lines in their optical spectra 
of \ion{He}{ii} and \ion{C}{iv}, \ion{He}{i}, and \ion{H}\ respectively.

Given the relatively long pulsation periods of white dwarfs, the realisation 
that white dwarfs are non-radial gravity-mode pulsators \citep{chan:72,wr:72} 
came soon after the discovery of the first variable white dwarf \citep{land:68}; 
that their photometric variations are primarily a manifestation of 
temperature perturbations rather than due to variations in geometry or surface 
gravity came only a decade later \citep{rkn:82}.

In order to subject the ZZ Cetis to asteroseismological analysis and
to provide constraints for pulsation models, it is crucial that the
eigenmodes associated with the observed periodicities be identified. 
Observationally, this means determining the value of the spherical degree, 
$\ell$, and azimuthal order, $m$. A third quantity, not an observable, is the 
radial order $n$; it specifies the number of nodes in the radial direction and 
can only be inferred by detailed comparison of observed mode periods 
with those predicted by pulsation models.

Mode identification for the ZZ Cetis is fraught with difficulties. In part,
the analysis has been hampered by an insufficient number of pulsation modes
excited to observable levels and mode variability over several different time 
scales. For the vast majority of ZZ Cetis, results have remained somewhat
ambiguous as the prerequisites for asteroseismological analysis were not met.
Even the Whole Earth Telescope (WET) campaigns \citep{nath:90} on several
objects \citep[e.g. G 117-B15A, ZZ Psc,][]{kepler:95,kman:98} were thwarted 
either by the small number of modes and lack of clear multiplet structure 
exhibited by these objects. Most efforts have focused on identifying similarities 
between the pulsational spectra of different stars to constrain the mode 
identification. This ultimately has a bearing not only on the determination of 
fundamental stellar properties but also on the mass of the superficial hydrogen 
layer \citep[e.g.][]{bradley:98}. It is clear that there is an acute need for 
more direct and complementary methods of pinning down the identification of the 
eigenmodes.

To this end, \citet{robetc:95} presented a method for inferring $\ell$ based on the
the wavelength-dependence of limb darkening, due to which observed mode amplitudes 
vary with wavelength in a manner that depends on $\ell$, but not on any of the other 
properties of the pulsation mode, such as $m$ or amplitude\footnote{At least for 
pulsation amplitudes of up to $\sim$5\% \citep{ik:01}.}. 
Thus, in a given star, modes having the same spherical degree will behave 
in the same manner. \citet{robetc:95} acquired photometric data in the ultraviolet 
of the ZZ Ceti star \object{G 117-B15A}. Application of Bayes' theorem and quantitative 
use of model wavelength-dependent pulsation amplitudes led them to infer that the 
largest amplitude mode of G 117-B15A had $\ell=1$. 

More recently, \citet{vkcw:00} and \citet{cvkw:00} used a variant of the above 
method to identify the spherical degree of the pulsation modes observed in ZZ Psc
(a.k.a. G 29-38) -- a star that has been notoriously erratic in the 
pulsation modes that it excites. Their investigation, which was based on amplitude 
changes within the Balmer lines at visual wavelengths {\em only\/}, yielded 
empirical differences between the modes that were best interpreted as several 
$\ell=1$ modes and one $\ell=2$ mode; the presence of modes of differing $\ell$ 
obviated the need for quantitative model comparisons making their identification 
more secure than any previous attempt. A surprising by-product of their analysis 
was the detection of variations in the line-of-sight velocity associated with the 
pulsations. Given the instrumentation available at that time, these were 
thought to be too small to measure \citep{rkn:82}.

The measurement of line-of-sight velocity variations associated with the
pulsations can be used in two complementary ways: (i) they provide a means 
with which to verify and constrain the theories of mode driving 
and, (ii) under certain theoretical assumptions, velocity variations in 
conjunction with flux changes provide an important new tool with which to 
probe the outer layers of pulsating white dwarfs.

In this study, we will only attempt to interpret our observations within 
the context of theories of mode driving via convection \citep{brick:83,brick:91,
gw:99a,gw:99b}, as opposed to theories which purport mode driving by
variants of the classical $\kappa$-mechanism \citep[e.g.][]{dzk:81,dv:81,wing:82}.
Unfortunately, as far as we are aware, testable predictions
for these models are not, as yet, available.

Within the context of the ``convective-driving'' picture the convection zone 
responds to perturbations from the adiabatic interior on time scales very much 
shorter ($\sim\!1$\,s) than the mode periods (hundreds of seconds). This allows 
the convection zone to absorb and release the flux perturbations cyclically, thus 
driving the pulsations. Combination frequencies (linear sums and differences of 
the real modes) arise naturally in the above picture and can, in principle, provide 
additional information. Furthermore, the relation between the flux and 
velocity variations allows one to estimate the total thermal capacity of the 
convection zone. Qualitative relations derived by \citet{gw:99a,gw:99b} show that
these are also sensitive to $\ell$. We explore these issues in Sect. \ref{sec:confront}.

Based on the success of identifying the $\ell$ index of the modes of ZZ Psc
using time-resolved spectroscopy, we present here, results from a similar analysis 
on HS 0507+0434B. Our primary aims are to attempt to determine the spherical 
degree ($\ell$) of the pulsations and to search for line-of-sight velocity variations 
associated with the pulsations. We also hope to constrain the properties 
of convection, a process which is poorly understood and therefore one of the main 
sources of uncertainty in the models.

We begin with a brief introduction to HS 0507+0434 in Sect. \ref{sec:hs} 
followed by a description of the data in Sect. \ref{sec:obs}. In Sects. 
\ref{sec:periodicities}, \ref{sec:velocities}, and \ref{sec:champ}, we attempt 
to extract the flux and velocity amplitudes and phases and the change in pulsation
amplitudes with wavelength. From Sect. \ref{sec:confront} onwards, we apply the 
constraints provided by all of the above to extricate quantities that allow us 
to both test and subsequently use the theory.
\section{HS 0507+0434}
\label{sec:hs}
HS 0507+0434, like many other white dwarfs, is the by-product of a 
survey for faint blue objects (the Hamburg Quasar Survey in this case).
It comprises two DA white dwarfs in a common proper-motion pair
which, as discussed by \citet{jord:98}, is of particular interest
for two reasons.
First, the effective temperatures of 20\,kK for HS 0507A and 11.7\,kK
for HS 0507B are such that the atmosphere of the former is completely
radiative, while the latter has an outer convection zone. Thus, one can
hope to use the parameters inferred from HS 0507A, which should be secure
-- since radiative atmospheres consisting of pure hydrogen are well understood --
to calibrate models for HS 0507B, which have to rely on uncertain models 
for convection. \citet{jord:98} found that only with relatively inefficient
versions of the Mixing Length Theory (MLT) were they able to find consistent 
solutions for both components in the system.
\begin{figure} 
\plotone{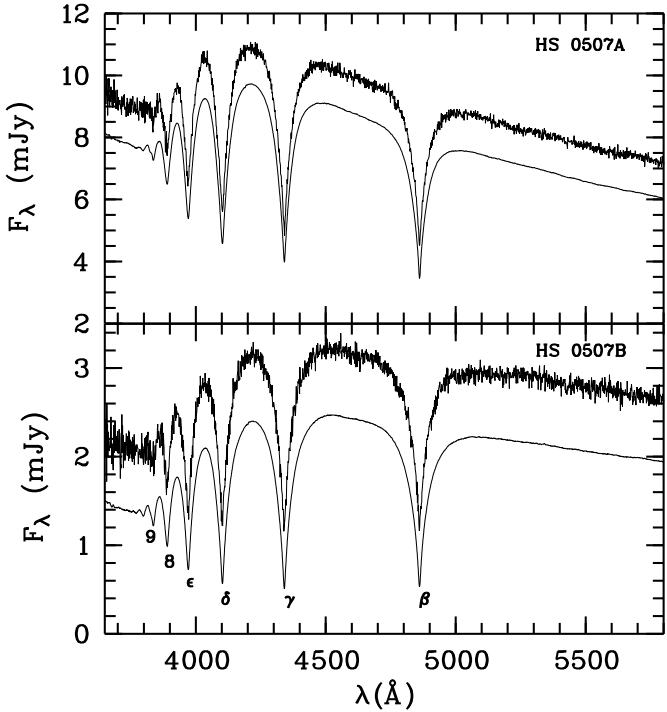}
\caption{Sample and average spectra of HS 0507A and B
in the 3650--5800\,{\AA} range with the
Balmer lines indicated. The average spectra are offset from the
sample spectra by 1\,mJy and 0.65\,mJy for the A and B components
respectively.}
\label{avspec}
\end{figure}

The second reason that HS 0507+0434 is of interest is that the temperature 
of HS 0507B -- as derived from its spectrum -- placed it squarely within 
the ZZ Ceti instability strip; fast photometry on this object revealed it, as 
expected, to be variable \citep{jord:98}.
A study of the temporal behaviour of HS 0507+0434B, based on photometric 
measurements collected over a total of seven consecutive nights, was recently
carried out by \citet{gh:00}. These authors were able to resolve a set
of three equally spaced triplets\footnote{Assuming spherical symmetry, frequencies 
of modes having the same $n$ and $\ell$ are degenerate. Slow rotation and/or 
a magnetic field lifts this degeneracy resulting in $2\ell+1$ and $\ell+1$ split
components respectively.}. Under the assumption that the splitting was due 
to the effect of slow rotation on $\ell=1$ modes, the observed triplets yielded 
an estimate of the rotation rate ($\sim\!1.5\,$days) and the $m$ values of 
the multiplets. 
They also found that in all three triplets, the $m=0$ component was much weaker 
than the $m=-1$ and $m=+1$ components, which were of roughly similar strength. 
Assuming that the intrinsic amplitudes were the same for all $m$ components, they 
estimated the inclination of the rotation axis with respect to the line of sight 
to be $\sim\!79\arcdeg$. The independent $\ell$ identifications afforded 
by clearly split multiplets are highly desirable as they can help to validate the 
identifications that rely on time-resolved spectroscopy only, given the 
differences between model spectra and observations.

As will become clear in the following sections, the advantage
of having a flux and velocity reference in the {\em same\/}
slit as the target greatly increases the accuracy of subsequent
measurements by making it possible to not only divide out atmospheric
fluctuations, but also to ensure that small random movements of the 
target in the slit are accounted for in the determination of the Doppler 
shifts of the Balmer-lines, thus making HS 0507+0434 an ideal
system on which to test theoretical predictions.
\begin{figure}
\plotone{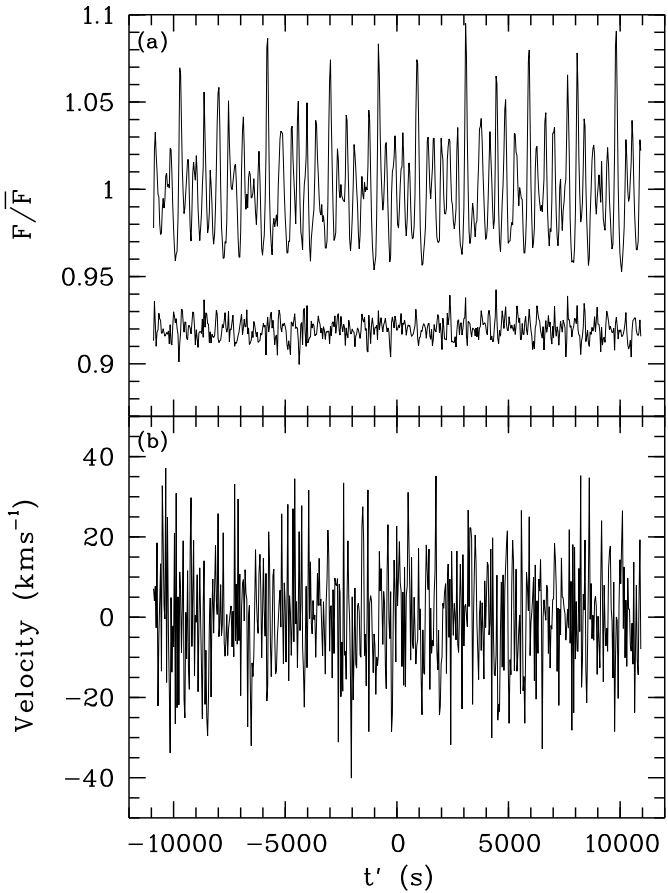}
\caption{Light and velocity curves of HS 0507B, shown here
for the first night only for the sake of clarity. t$'$ = t$-$10:15:14
U.T. i.e. relative to the midpoint of the time series.
\textbf{(a)} Fractional variations in the continuum region between
$\sim$ 5300 -- 5800\,{\AA}. The variations are with respect to the
continuum of HS 0507A in the same wavelength region.
Note that the maxima are stronger and sharper than the minima
as is typical for ZZ Ceti type pulsators.
The bottom curve shows the residuals (offset by $+$0.92)
after fitting 26 sinusoids with amplitudes larger than
0.15\% (see Table \ref{hstab});
\textbf{(b)} variations in the line-of-sight velocity also with
respect to HS 0507A -- as derived from fitting profiles
to H$\beta$, H$\gamma$, and H$\delta$. No obvious variations
are present and indeed we detect only marginal velocity variations
associated with the pulsation modes (Sect. \ref{sec:velocities}).}
\label{ndtligovel}
\end{figure}

\section{Observations and Data reduction}
\label{sec:obs}
\noindent
HS 0507+0434 was observed on the nights of $10^{\mathrm{th}}$ and
$11^{\mathrm{th}}$ December 1997 using the Low Resolution Imaging
Spectrometer (LRIS) on the Keck II telescope \citep{oke:95}.
An 8$\farcs$7-wide slit was used together with a 600 line\,mm$^{-1}$  
grating covering the 3450-5960\,{\AA} range at 1.25\,\AA\,pixel$^{-1}$. 
On-chip binning by a factor of two was implemented in the spatial direction.
The first night was photometric and the seeing of 1$\farcs$2 set the wavelength
resolution to 7\,{\AA}. There were thin patches of cirrus on the second
night; furthermore, there was a great deal of windshake, which resulted in 
a worse resolution (9\,{\AA}) although the situation improved towards the
end of the run as the elevation decreased.
A contiguous set of 560 24\,s exposures were acquired from 7:13:35 to
13:17:17 U.T. on the first night and a further 280 exposures were
obtained from 7:10:50 to 10:12:24 U.T. on the following
night. For both nights, the series was preceded by exposures
of the flux standard G 191-B2B (five and ten integrations of 4\,s
respectively), followed by HgKr arc spectra for wavelength calibration,
and halogen flat-field frames.
A total of 64 frames from the second night had to be discarded
due to a malfunction of the guider, which resulted in the targets
drifting off the slit.
We decided to fix the slit at the position angle required to have both
components of the system in the slit and deal with the effects of
differential refraction during the reduction of the data.
The decision to use a wide slit stemmed from the need to acquire 
`photometric' measurements of HS 0507 A and B in addition to measurements
of variations in the line-of-sight velocity (see Sect. \ref{sec:periodicities}).

The reduction of the data was carried out using 
MIDAS\footnote{The Munich Image Data Analysis System, 
developed and maintained by the European Southern Observatory.}
and using routines running in the MIDAS environment written specifically
for the LRIS instrument and this data set.
In detail, our reduction procedure entailed: 
(i) subtracting the CCD bias level of each amplifier separately using
the overscan region; 
(ii) correcting for the (small) non-linearities that most CCDs are
prone to, using linearity curves of the CCD derived from previous
observations using the LRIS instrument;
(iii) correcting for the gain difference between the 2 amplifiers as
derived from halogen lamp frames;
(iv) flat-fielding using an average of the halogen frames for the
first night and an average of dome-flats for the second night (as 
these gave smoother results); 
(v) sky subtraction;
(vi) correcting for the error introduced by dividing by a flat field
taken through slightly non-parallel slits (described in more detail
below);
(vii) optimal extraction of the spectra using a method akin to that
of Horne (1986);
(viii) wavelength calibration using Hg/Kr frames (by tabulating
wavelengths for each pixel rather than by rebinning the spectra), with
an offset determined from star A, and including a correction for
refraction (see below); and
(ix) flux calibration with respect to the
flux standard G 191-B2B, using the model fluxes of \citet{modflux:95}
and using the extinction curve of \citet{ext:88} to
correct for small differences in airmass.

Among the usual preprocessing stages described above, two additional
non-standard steps were required given our use of a wide slit, namely the 
need to correct for non-parallel slit-jaws and the need to account for 
the effects of random stellar wander in the slit due, for instance, to 
seeing-related changes, windshake, and tracking uncertainties.

\begin{figure*}
\plotone{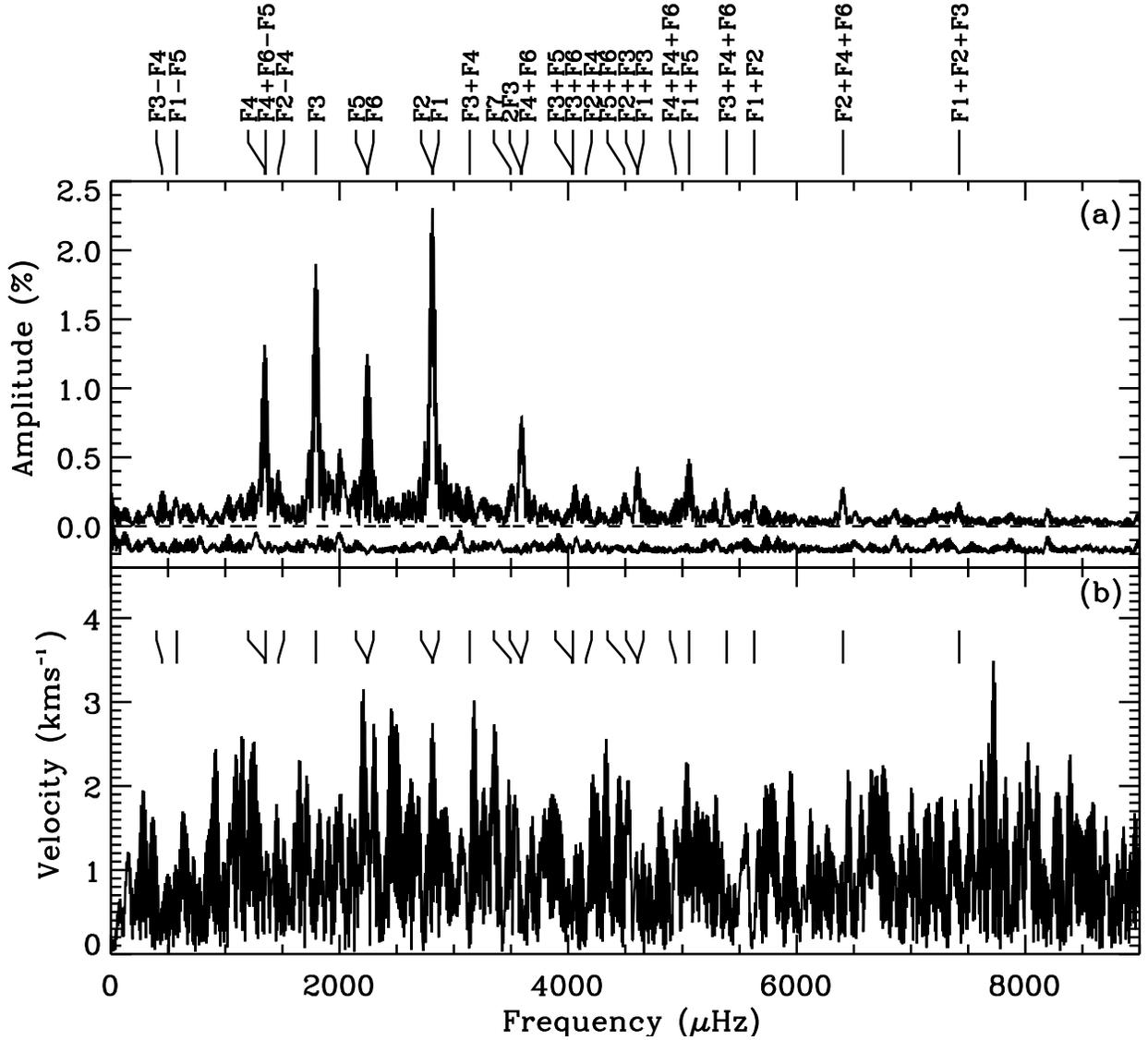}
\caption{\textbf{(a)} Fourier transform of the light curve (top) and residuals 
(bottom) offset by $-$0.2\%. The maximum amplitude of peaks longward of 
9000$\mu$Hz is $\sim$ 0.09\%. \textbf{(b)} Fourier transform of the velocity 
curve. The frequencies inferred from the light curve are indicated. 
None of the frequencies present in the light curve have obviously
significant associated velocity modulation. This is confirmed by our
detailed analysis (Sect. \ref{sec:velocities}), with the possible exception 
of F2, for which we find that associated signal has a probability of only 
2\% of being due to chance, constituting a marginal detection. (Note that
without prior information from the light curve regarding which
frequencies to expect, this peak would also be insignificant).
Both (a) and (b) have been constructed using data from both nights.}
\label{hsflxovel}
\end{figure*}

The former correction is necessitated by the fact that while the flat-fields
and sky background are influenced by the shape of the slit, the light from
the targets is not, as no light falls outside the slit jaws. The sky background
was determined by fitting first degree polynomials for each step along the 
dispersion direction excluding two regions of 26 and 30 binned pixels 
corresponding to HS 0507A and HS 0507B respectively, and multiplying the result 
by a function describing the shape of the slit. Finally, we divided the images 
by an average flat field frame that was normalised along the spatial direction 
in order to maintain the correct relative flux ratio between the two target 
objects.

A second non-standard correction is required to purge the data of any 
extraneous effect that introduces shifts in the positions of the Balmer
lines, such as instrumental flexure, refraction, and wander. In principle,
all these changes should affect HS 0507A and B in the same way, and one could 
simply measure wavelengths relative to HS 0507A only. However, we prefer the 
following methodology, as it not only permits us to assess the quality that one 
may expect in the absence of such a fortuitous local calibrator, but it also 
permits us to verify the stability of HS 0507A and provides insight 
into any systematic effects that may have crept into our analysis of other 
targets. We first correct the spectra of both stars for instrumental flexure 
and refraction\footnote{See Appendix \ref{sec:acheck} where we scrutinize HS 0507A 
for any signs of variability.} and only then apply a correction for stellar 
wander, using the measured offsets of H$\gamma$ of HS 0507A as fiducial points. 
Having thus corrected each pixel in the spectrum for the effects of instrumental 
flexure, differential refraction, and random wander in the slit, we redid the
wavelength and flux calibrations. Any remaining offsets of the Balmer
lines from their respective laboratory wavelengths must now be due to
intrinsic variations only.

An estimate of the remaining scatter due to wander in the derived velocities can 
be obtained from the shifts in the slit of the spatial profiles of both HS 0507A 
and HS 0507B and assuming that the scatter in these shifts is also representative 
of the scatter in the dispersion direction. Having fitted Gaussian functions to 
the spatial profiles, we take the standard deviation of the difference of the 
resulting centroid positions of HS 0507B with respect to those of HS 0507A as 
an indicator of the scatter and obtain a value of $\sim$2\kms at H$\gamma$ as 
an average value for both nights. This is well below the measurement errors 
(see Sect. \ref{sec:velocities}). Note that the root-mean-square scatter for ZZ 
Psc was 14\,\kms at H$\gamma$ \citep{vkcw:00}. The value of having a reference 
object in the slit thus cannot be overemphasized.

The final average spectra of HS 0507+0434 A and B, as well as
sample individual spectra, are shown in Fig. \ref{avspec}.

\begin{table*}[t]
\caption[]{Pulsation frequencies and other derived quantities from the light and velocity curves 
of HS 0507B}
\fontsize{7.5}{10}\selectfont
 \begin{centering}
  \begin{tabular}{c@{\hspace{0.2cm}}lcccccccc}
 \label{hstab}
       &          &                 &              &            &                &       &         &             \\
\hline 
&  Mode &   Period &  Frequency      & $\Phi_{L}$   &   $A_{L}$ &  $A_{V}$       & $\Phi_{V}$         & $R_{V}$      & $\Delta\Phi_{V}$ \\
&       &    (s)   & ($\mu$Hz)      & (\arcdeg)    &  (\%)     &  (\kms)  & (\arcdeg)          & (Mm rad$^{-1})$ & $(\arcdeg)$ \\
\hline
\sidehead{Real Modes:}                
 \multirow{2}{-0.2cm}{\{}
& F1 ($f_{10}$) $m$ =\hspace{1.7pt}\phs1& 354.9 $\pm$ 0.03 &2817.5 $\pm$ 0.3 &\phn$-$27 $\pm$ 4 & 0.68 $\pm$ 0.11 &2.0 $\pm$ 1.0 &  \phn\phs61 $\pm$ 29 & 17 $\pm$ 9 &\phs\phn88 $\pm$ 29    \\
& F2 ($f_8$) \phn$m=-1$ & 355.8 $\pm$ 0.02 &2810.9 $\pm$ 0.2 &\phn$-$57 $\pm$ 1 & 2.27 $\pm$ 0.06 &2.6 $\pm$ 1.0 & \phn$-$13 $\pm$ 23 & \phn6 $\pm$ 3 & \phs\phn44 $\pm$ 23\\
& F3 ($f_2$) \phn$m=-1$& 557.7 $\pm$ 0.03 &1793.2 $\pm$ 0.1 &\phn\phn$-$6 $\pm$ 1 & 1.87 $\pm$ 0.04 &0.8 $\pm$ 0.8 & \nodata & \phn4 $\pm$ 4  & \nodata  \\	
& F4 ($f_1$) & 743.0 $\pm$ 0.13 &1346.1 $\pm$ 0.2 &$-$153 $\pm$ 2 & 1.39 $\pm$ 0.04 &0.3 $\pm$ 1.1 & \nodata  & \phn2 $\pm$ 9 & \nodata  \\
 \multirow{2}{-0.2cm}{\{}
& F5 ($f_5$) \phn$m=-1$& 446.2 $\pm$ 0.05 &2241.1 $\pm$ 0.2 &\phn$-$0.2 $\pm$ 3 & 1.10 $\pm$ 0.09 &1.6 $\pm$ 1.1 & \nodata & 10 $\pm$ 7 & \nodata \\
& F6 ($f_7$) \phn$m$ =\hspace{1.5pt}\phs1& 444.8 $\pm$ 0.06 &2248.4 $\pm$ 0.3 &\phn\phs74 $\pm$ 3 & 1.36 $\pm$ 0.07 &2.3 $\pm$ 1.1 &  \phn\phs79 $\pm$ 26 & 12 $\pm$ 5  & \phs\phn\phn4 $\pm$ 27    \\	   
& F7  & 286.1 $\pm$ 0.04 &3495.8 $\pm$ 0.5 & \phs147 $\pm$ 6 & 0.36 $\pm$ 0.04 &0.8 $\pm$ 0.8 &  \nodata & \phn9 $\pm$ 9 & \nodata \\

\sidehead{Combination Modes:}
&  Mode &   Period &  Frequency      & $\Phi_{L}$   &   $A_{L}$ &  $A_{V}$       & $\Phi_{V}$ & $R_{C}$ & $\Delta\Phi_{C}$ \\
&       &    (s)   & ($\mu$Hz)  & (\arcdeg) &  (\%)  &  (\kms) &(\arcdeg) &  & (\arcdeg)\\ 
        &          &                 &              &            &                &       &         &     \\
& F4+F6 & \phn278.2 & 3594.4  & $-131 \pm \phn4$ & 0.80 $\pm$ 0.05 & 0.9 $\pm$ 1.1 & \nodata & 20.7 $\pm$ 1.8 & \phn$-53 \pm \phn5$ \\
& F1+F3 & \phn216.9  &4610.7  &\phn$-$32 $\pm$ \phn7 & 0.38 $\pm$ 0.04 & 0.2 $\pm$ 1.0 & \nodata & 15.0 $\pm$ 3.0 & \phn\phn\phs1 $\pm$ \phn8\\
& F2+F3 & \phn217.2  &4604.2  &\phn$-$73 $\pm$ \phn8  & 0.30 $\pm$ 0.05 & 1.1 $\pm$ 1.0 & \nodata  & \phn3.5 $\pm$ 0.6 & \phn$-$10 $\pm$ \phn8\\
& F2+F4 & \phn240.6 &4156.8  & \phs129 $\pm$ 11 & 0.20 $\pm$ 0.04 & 0.4 $\pm$ 0.8 & \nodata & \phn3.0 $\pm$ 0.6 & \phn$-$21 $\pm$ 12\\
& F3+F6 & \phn247.4 & 4041.7  & \phn\phs71 $\pm$ \phn8 & 0.33 $\pm$ 0.04 & 1.0 $\pm$ 1.0 & \nodata  & \phn6.4 $\pm$ 0.9 & \phn\phn\phs2 $\pm$ \phn8\\
& F3+F5 & \phn247.9 & 4034.4  &\phn$-$57 $\pm$ 17 & 0.16 $\pm$ 0.04 & 0.4 $\pm$ 1.1 & \nodata & \phn3.9 $\pm$ 1.1 & \phn$-$51 $\pm$ 17\\
& F1+F5 [F2+F6] & \phn197.7  &5058.6  & \phn$-$12 $\pm$ \phn4 & 0.50 $\pm$ 0.04 & 1.4 $\pm$ 0.8 & \nodata & 32.8 $\pm$ 6.2 & \phn\phs15 $\pm$ \phn7\\
& F2$-$F4 &\phn682.6 &1464.9 &\phn\phs41 $\pm$ \phn6  & 0.37 $\pm$ 0.04 & 1.3 $\pm$ 0.8 & \nodata & \phn5.9 $\pm$ 0.6 & \phn$-$55 $\pm$ \phn6\\
& F5+F6 &\phn222.7  &4489.6  &\phn\phs80 $\pm$ \phn8  & 0.27 $\pm$ 0.04 & 0.8 $\pm$ 0.8 & \nodata & \phn9.0 $\pm$ 1.5 & \phn\phn\phs5 $\pm$ \phn9\\
& F1+F2 &\phn177.7 &5628.4 & $-$140 $\pm$ 10  & 0.21 $\pm$ 0.04 & 0.5 $\pm$ 0.8 & \nodata  &\phn6.7 $\pm$ 1.6 & \phn$-$57 $\pm$ 11\\
& F3$-$F4 [F5$-$F3] &  2236.5 &\phn447.3 & $-$102 $\pm$ \phn9  & 0.23 $\pm$ 0.04 & 0.3 $\pm$ 0.8 & \nodata &\phn4.3 $\pm$ 0.7 & \phs111 $\pm$ 10\\
& F1$-$F5 &  1735.0 &\phn576.4  & \phs144 $\pm$ 10  & 0.22 $\pm$ 0.04 & 0.9 $\pm$ 0.8 & \nodata & 14.4 $\pm$ 3.5 & \phs170 $\pm$ 11\\
& F3+F4 & \phn318.6 &3139.2 & \phs160 $\pm$ 11 & 0.19 $\pm$ 0.04 & 1.0 $\pm$ 0.8 & \nodata &\phn3.6 $\pm$ 0.7 & \phn$-$41 $\pm$ 11\\
& 2F3 [F4+F5] & \phn278.8 &3586.5 & $-$128 $\pm$ 17  & 0.19 $\pm$ 0.05 & 1.4 $\pm$ 1.0 & \nodata & \phn5.4 $\pm$ 1.5 & $-$116 $\pm$ 17\\
& F1+F2+F3  & \phn134.7 &7421.6 & $-$128 $\pm$ 14 & 0.16 $\pm$ 0.04 & 0.3 $\pm$ 0.8 & \nodata & \phn90 $\pm$ 30 & \phn$-$38 $\pm$ 14\\ 
& F2+F4+F6 [2F3+F1]& \phn156.1 &6405.2 & \phs151 $\pm$ \phn7 & 0.28 $\pm$ 0.04 & 0.7 $\pm$ 0.8 &  \nodata & 108 $\pm$ 16 & \phn$-$74 $\pm$ \phn8\\
& F3+F4+F6 &\phn185.6 &5387.6 & \phs152 $\pm$ \phn8  & 0.25 $\pm$ 0.04 & 0.7 $\pm$ 0.8 & \nodata & 117 $\pm$ 19 & $-$124 $\pm$ \phn9\\
& 2F4+F6 &\phn202.4 &4940.3 & \phn$-$14 $\pm$ 13 & 0.17 $\pm$ 0.04 & 1.8 $\pm$ 0.8 & \nodata & 208 $\pm$ 50 & $-$143 $\pm$ 14\\
& F4+F6$-$F5 [$m$=1, F4]& \phn739.0 &1353.3 & \phn\phn\phs1 $\pm$ 18 & 0.24 $\pm$ 0.05 & 1.3 $\pm$ 1.0 & \nodata  & 190 $\pm$ 43 & \phs\phn79 $\pm$ 18\\
\hline
\end{tabular}
\tablecomments{The frequencies listed in brackets next to the real modes refer to the 
nomenclature preferred by \citet{gh:00}. The curly brackets next to two pairs of real 
modes indicate that these are multiplets sharing the same $\ell$ but differing $m$ 
values (see also Sect. \ref{sec:combfreq}). We have followed the convention that positive 
values of $m$ correspond to prograde modes. $R_{V} = A_{V}/(2\pi fA_{L})$, in units of
Mm rad$^{-1}$, has the physically intuitive meaning of being the typical distance a fluid 
element travels on the surface for a pulsation with a fractional amplitude of unity.
$\Phi_L$ and $\Phi_V$ are the phases of the sinusoids fit to the light and
velocity curves respectively with $\Delta\Phi_{V} = \Phi_{V} - \Phi_{L}.$
For the combination frequencies, $R_{C} = A_{L}^{i\pm j}/(n_{ij}A_{L}^{i}A_{L}^{j})$ 
where $n_{ij}$ is unity for $i=j$, 2 for 2-mode combinations etc. and 3 for
3-mode combinations involving the first harmonic, e.g. 2F4+F6.
$\Delta\Phi_{C} = \Phi_{L}^{i\pm j} - (\Phi_{L}^{i} \pm \Phi_{L}^{j}$).
Potentially misidentified modes are listed in square brackets next to the combination 
modes.  Note that this list is not an exhaustive one. The frequency of the combination 
F4+F6$-$F5 is at the location of the $m=1$ multiplet c.f. F4 (we infer in Sect. 
\ref{sec:champ} that F4 has $\ell=1$ and in Sect. \ref{sec:combfreq} that it probably 
has $m=-1$). }
\end{centering}
   \end{table*}
\section{Periodicities in the Light Curve} 
\label{sec:periodicities}
Continuum light curves for each night were constructed by dividing the average
flux in the 5300--5800\,{\AA} region of HS 0507B by that of HS 0507A (see Figs. 
\ref{avspec}, \ref{ndtligovel}a). Note that the time axis has been computed
relative to the middle of the time series in order to minimize any covariance
between the frequencies and phases of the sinusoids we subsequently use to
fit the light curve ($t' = t-16:33:20$ U.T.).
The Fourier transform of the light curve obtained by merging the data from 
both nights is shown in Fig. \ref{hsflxovel}a and is typical of that of the
ZZ Cetis.

We determined the periodicities consecutively in terms of decreasing amplitude 
by means of an iterative process: using an approximate value for the 
frequency of the highest peak, we fitted a function of the form 
$A\cos(2\pi ft'-\phi) + C$ where $A$ is the amplitude, $f$ the frequency, 
$\phi$ the phase and $C$, a constant offset. The fit yielded $A$, $f$ and 
$\phi$. The process was repeated, adding a new sinusoid each time, and fitting 
for all parameters simultaneously until no peaks with amplitudes $\simgt 0.15\%$ 
could be identified in the Fourier transform of the residuals.

Combination frequencies were identified by searching for linear combinations
of the real modes and these were fitted by fixing the frequencies to those of 
the corresponding combination of the real mode frequencies. We imposed the 
requirement that the combination frequency in question have the smallest 
amplitude of the frequencies involved. Identification of combination frequencies 
was not always easy due to degenerate combinations (e.g. F3$-$F4 $\simeq$ F5$-$F3
given our resolution). In such cases, we opted for the largest amplitude of the 
combination frequency and the lowest possible error yielded by the fit. The light 
curve is not free of periodicities after having subtracted the frequencies listed 
in Table \ref{hstab}; although we continued to fit further combination
modes, we found that the choice of combinations became rather arbitrary and
hence we do not report these here.

Two statements can already be made on the basis of Table \ref{hstab}. 
First, as F1 and F2 have the same splitting as the F5 and F6 multiplet components, 
they therefore very probably also share the same $\ell$ value. From the analysis of 
\citet{gh:00}, who detect triplets in each of these groups, they are likely to have 
$\ell=1$ and $m=\pm1$. Second, the periods of the real modes are very similar to 
those observed by \citet{kman:98} in a time series spanning 10 years for ZZ Psc, a 
similar, but slightly cooler white dwarf. In that star, these authors find real 
modes at periods ranging from 110 - 915\,s. For $\ell=1$, these correspond to 
successive radial orders ($n$) from 1 to 18. The modes in common with HS 0507B are 
at 284, 355, and 730\,s, which they identify as having radial orders of 4, 5, and 
13 respectively. 
The $n$ = 7 mode in their model is the only mode (from $n$ = 1 - 18) 
not detected in the ZZ Psc data. Interestingly, a mode at the expected
frequency is present in HS 0507B as the F5,F6 multiplet pair (445\,s) at 
the expected separation for $\ell=1$ modes. It is tempting to conclude that,
as for the hot ZZ Cetis \citep{clem:93}, the cool ZZ Cetis have remarkably 
similar structure. One should bear in mind, however, that similar period 
structures may well be produced for a number of different combinations of 
mass and hydrogen-layer thickness \citep[e.g.][]{bradley:98}.
 
\begin{figure}
\plotone{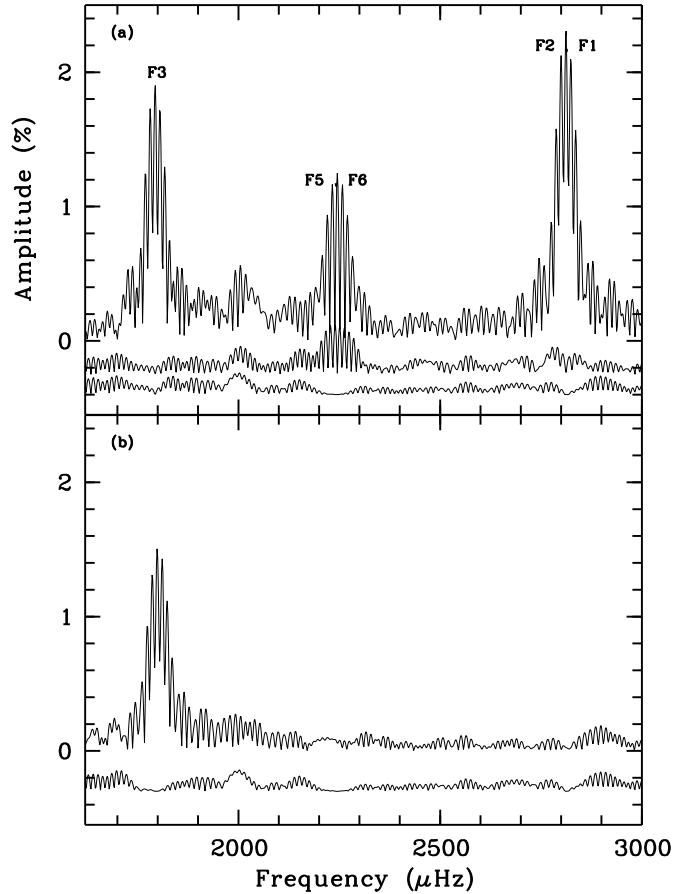}
\caption{\textbf{(a)} The top curve shows the FT of the light curve in the
$1600-3000\,\mu$Hz range. The middle curve shows the FT of the
residuals after subtracting sinusoids with frequencies corresponding
to all real modes except F1 and F6, which are multiplets in the same
period group as F2 and F5 respectively, while the bottom curve shows
the effect of including F1 and F6 in the fit. These two lower curves
have been offset by $-$0.25\% and $-$0.40\% respectively.
While there are some feeble signs of residual modulation around F1 and F2,
this is clearly not the case for F5 and F6. Fitting an additional sinusoid
near the F1,F2 pair in the hope of recovering the missing component is of no
avail. \textbf{(b)} Top curve: residuals in the FT after subtracting three 
sets of triplets with frequencies and amplitudes fixed to those listed in 
\citet{gh:00}. The large residual corresponds to the $m=1$ mode at 1800.7$\mu$Hz 
showing that the amplitude of this mode has changed considerably since our 
observations, from 1.7 to $\lesssim 0.5$\%. The bottom curve (offset by $-$0.3\%) 
shows the effect of leaving the amplitudes free to vary while leaving the 
frequencies fixed.}
\label{multipdft}
\end{figure}

We refrain from a detailed comparison with the study of \citet{gh:00} given 
our lower frequency resolution, but a few points are noteworthy. While the 
frequencies of the modes detected in our data are entirely consistent 
with those found by \cite{gh:00}, our lower resolution does not permit us to 
confirm the presence of the $m=0$ multiplet components independently (see Fig. 
\ref{multipdft}a); these were also the weakest modes in all triplets and thus 
should not greatly influence the amplitudes of the $m=\pm1$ components. As a check, 
we attempted to recover the real modes identified by \citet{gh:00} that were not 
detected by us, by first imposing the amplitudes and frequencies listed by 
\citet{gh:00} and subsequently leaving the amplitudes free to vary. The residuals 
are shown in Fig. \ref{multipdft}b. We find that our data are consistent with all 
but one of the components and have amplitudes consistent with those derived by 
\citet{gh:00}. The only exception is the $m=1$ mode at 1800.7\,$\mu$Hz (in the 
same multiplet as our F3), for which \citet{gh:00} found an amplitude of 1.7\%, 
while it has an amplitude of only $\la\!0.5\%$ in our data and is not even 
required in the fit. For completeness, we note that the amplitudes and phases 
we derive using the frequencies of \cite{gh:00} are entirely consistent with 
those listed in Table \ref{hstab}.

\section{Periodicities in the line-of-sight velocities}
\label{sec:velocities}
Variations in the equivalent widths and profiles of the Balmer
lines reflect temperature and line-of-sight velocity variations.
The details of the complex interactions between flux and velocity 
that result in the observed spectral line are not well-understood,
but are likely to be separable to first order.

In order to search for line-of-sight velocity variations, we 
determined Doppler shifts for the Balmer lines by fitting
a combination of a Gaussian and a Lorentzian profile which
provided a good fit to the lines; the continuum was 
approximated by a line.
We chose the wavelength intervals 4660$-$5088\,{\AA}, 4214$-$4514\,{\AA},
and 4028$-$4192\,{\AA} for H$\beta$, H$\gamma$, and H$\delta$ respectively
to carry out the fits and insisted that the central wavelengths of the
Gaussian and Lorentzian functions be the same.
We repeated the above process for H$\gamma$ only, fitting the logarithm of
the flux instead as a check of our (additive) fitting method. The velocities
thus derived were nearly identical with those obtained by fitting the
flux only.
Possible systematic effects arising from different methods of deriving
these velocities are discussed in detail in \citet{vkcw:00}. 
The resulting velocity curve is shown in Fig. \ref{ndtligovel}b, 
while the associated Fourier transform is shown in Fig. \ref{hsflxovel}b.
The typical uncertainty in a single measurement for HS 0507B is 14\,\kms,
much larger than those due to differential wander between HS 0507A and B 
which was estimated to be $\simlt$ 2\,\kms (Sect. \ref{sec:obs}).

Simple checks confirm that the errors in the measurement of the
line-centroids of each of the three Balmer lines are uncorrelated.
For instance, the standard deviation of the average velocity as
computed from an average of all three lines decreases by a factor
$(\sigma_\beta + \sigma_\gamma + \sigma_\delta)^{\frac{1}{2}}$/3
compared to that associated with a single line, as would be expected 
if the errors were mutually independent (here, $\sigma_{\beta,\gamma,\delta}$ 
are the respective standard deviations in the velocities as derived 
from H$\beta$, H$\gamma$, and H$\delta$).
The velocity differences between the lines also offer an independent
estimate of the measurement uncertainty; e.g. one expects the ratio
$(v_{\beta} - v_{\gamma})/(\sigma_\beta^2 + \sigma_\gamma^2)^{\frac{1}{2}}$
to have a mean of zero and a standard deviation of unity if the errors 
associated with each line are uncorrelated. Our values of the standard 
deviation of the above ratio for the two nights are 1.2 and 1.3, i.e., 
roughly consistent with unity. We deem our error estimates to be credible.

The Fourier Transform of the velocity curve (Fig. \ref{hsflxovel}b)
is striking in that it shows no strong peak at any frequency, not even
at the frequencies corresponding to the dominant variation in the light 
curve. We will explore the possible cause of this result in later sections.
We can, nevertheless, place interesting upper limits on the velocity to flux
ratio of each observed mode. The motivation for doing so is that the velocity
to flux ratios ($R_V$) have, to date, only been determined for one star 
\citep{vkcw:00}; yet, these are essential for comparison with theoretical 
predictions. While the noise level prevents the velocity curve from being
used in the absence of external information, the light curve provides 
just such external, independent, information as to the periodicities we expect 
to find in the velocity curve. We can exploit this additional information by 
imposing the frequencies we find in the light curve on the velocity curve and 
measuring the velocity amplitudes at these pre-specified frequencies.  
We can subsequently ask if a velocity peak at a known frequency is 
signficant. We detail this procedure below.

As just described, we looked for modulations in the velocity curve by fitting 
the velocity curve with sinusoids, the frequencies of which were fixed at those 
obtained from the light curve. As with the light curve, the calibration relative 
to HS 0507A removed all slow variations and only a constant offset was 
additionally included in the fit. The resulting velocity amplitudes and phases are 
listed in Table \ref{hstab}. We find marginally significant velocity variations at
F1, F2, and F6. If these are real, it is surprising that we do not find significant
velocity variations at F3 and F4, in spite of these modes having stronger 
flux modulations than F1 and F6. For F4, this is due in part to the proximity of 
the combination mode F4+F6$-$F5. Excluding all combination modes from the fit 
(theoretically, these are not expected to have associated velocity variations) 
results in an increase of the velocity amplitude of F4 to 0.8\,\kms\ ($R_V = 7\pm7$) 
while the amplitudes for the other real modes change by less than 0.2\,\kms. 

We carried out Monte Carlo simulations in order to ascertain the likelihood
that the peaks as large as those we detected might occur simply by chance.
Our simulations were conducted thus: we randomly distributed the velocities
with respect to the times and then fit these velocity curves in exactly
the same way as the observations i.e., by fixing the frequency of the sinusoids 
to those derived from fits to the light curve as described above. This procedure 
was repeated 1000 times and the number of peaks with amplitudes larger than those 
measured at the frequencies corresponding to the real modes were counted. 
The results corroborated our error estimates: the modulation at F2 was
most significant, having a probability of only 2\% of being a random 
occurrence, while the modulations at F1 and F6 had probabilities of 9\% and 4\% 
respectively of being chance occurrences. In what follows, we treat 
these measurements as upper limits.

In summary, we find that at the frequency of F2, the mode with 
the largest photometric amplitude, there is evidence for an associated
velocity signal with an amplitude of $2.6\pm1.0{\rm\,km\,s^{-1}}$,
which our Monte-Carlo simulation suggests has a probability of only
2\% of being due to chance. If taken at face value, we have thus
detected velocity variations in a second ZZ Ceti type pulsator.  A
more conservative conclusion is that the upper limit to the velocity
variations is $3{\rm\,km\,s^{-1}}$ (at 99\% confidence).
Before addressing what one can infer from the (limits to) the amplitudes, 
we first present a discussion of the chromatic amplitudes, from which 
we attempt to infer the spherical degree of the real modes.
\section{Chromatic Amplitudes}
\label{sec:champ}
\begin{figure}[!t]
\plotone{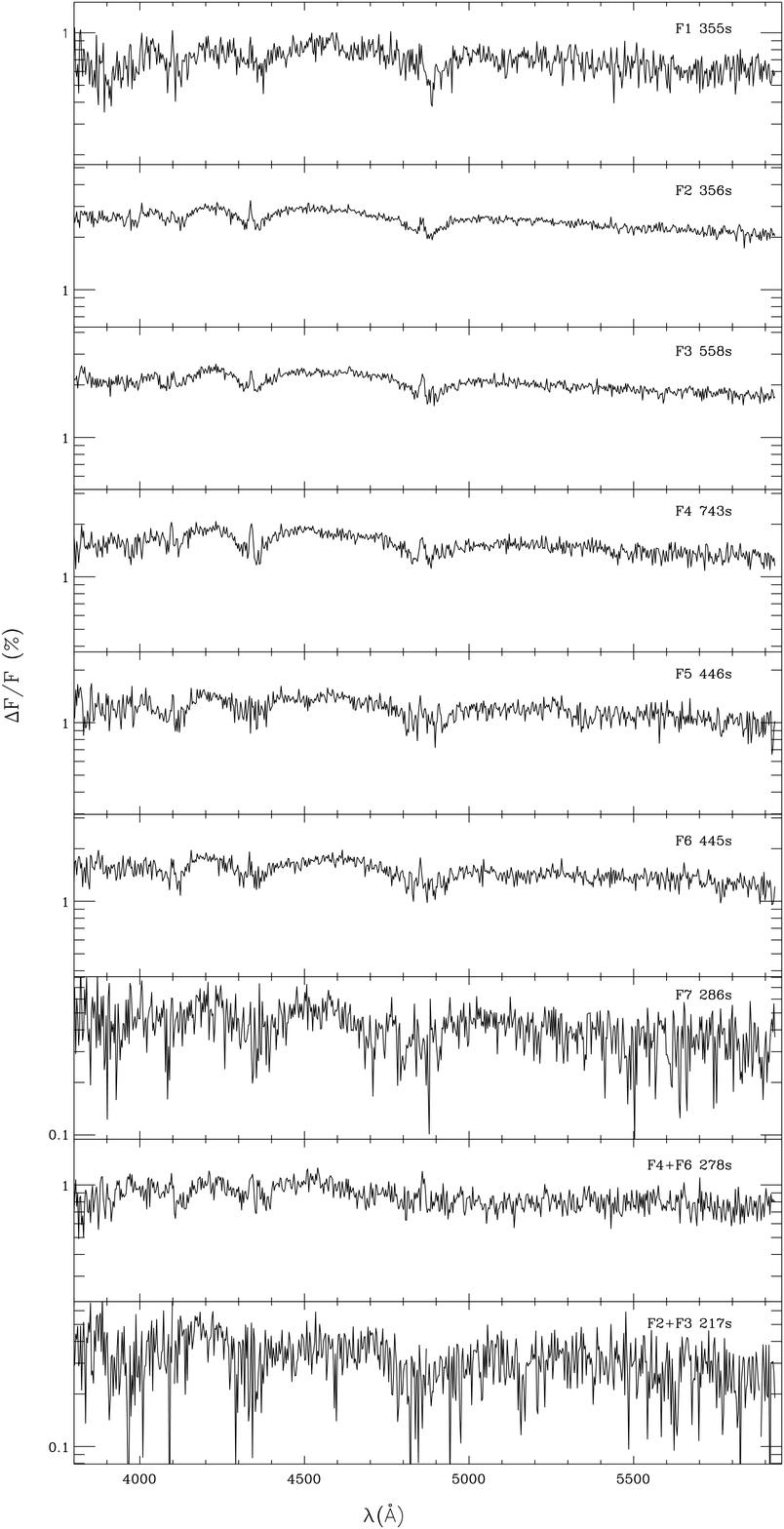} 
\caption{Chromatic amplitudes for the 7 real modes and 2 of the
stronger combination modes. The period is indicated next to the name.
From photometry, we know that F1 and F2 are modes having the same
$\ell$ but different $m$ values; likewise for F5 and F6. Note the shape and
curvature of the continuum between the line-cores. The chromatic amplitudes
confirm that at least the stronger modes share the same $\ell$ value (see also
Fig. \ref{fig:mixmod}).}
\label{fig:hschamp}
\end{figure}
As was noted by \citet{robetc:95}, the wavelength dependence of the fractional 
pulsational amplitude, $\Delta\,F_{\lambda}/F_{\lambda}$, can potentially provide
a clean measurement of the spherical degree ($\ell$) of the modes as it is not 
only independent of the inclination and $m$ (and therefore also to unresolved 
modes having the same $\ell$ but different $m$, but is also insensitive
to the details of flux calibration.\footnote{Note that the inference of 
$\ell$ from the observed chromatic amplitudes {\em only\/}, can, in principle,
be made {\em without\/} even the qualitative use of models. Although the use of 
models necessarily brings with it a dependency on the assumed convective 
efficiency, the qualitative use of model chromatic amplitudes computed with 
differing mixing length prescriptions does not affect the mode identification. 
The $\ell=2$ mode in \citet{cvkw:00} is always better matched with an $\ell=2$ 
model regardless of the mixing length prescription.}

We have computed these fractional pulsation amplitudes at each wavelength
(hence the name, `chromatic amplitudes') by fitting the real and combination 
modes listed in Table \ref{hstab} in each 3\,{\AA} bin, where the choice of
bin size reflects a compromise between adequate signal to noise and resolution.
The frequencies of the modes were fixed at the values shown in Table \ref{hstab} 
while the amplitudes of the modes were left free to vary. At first,
we also left the phases free, but found that these did not show any
significant signal. Since \cite{cvkw:00} found from their high signal-to-noise 
ratio data that the phases show very little variation\footnote{Note that
\cite{cvkw:00} did find phase changes across the Balmer lines, but at levels 
too low for us to detect. They also found a slight, unexplained slope of phase 
as a function of wavelength. We also find a similar variation (2-3\arcdeg) for the 
strongest modes.}, we decided to determine the amplitudes with the phases fixed 
to the values listed in Table \ref{hstab}. The resulting chromatic amplitudes are
shown in Fig. \ref{fig:hschamp} (allowing the phases to vary produces a very
similar figure).

\begin{figure}[!t]
\plotone{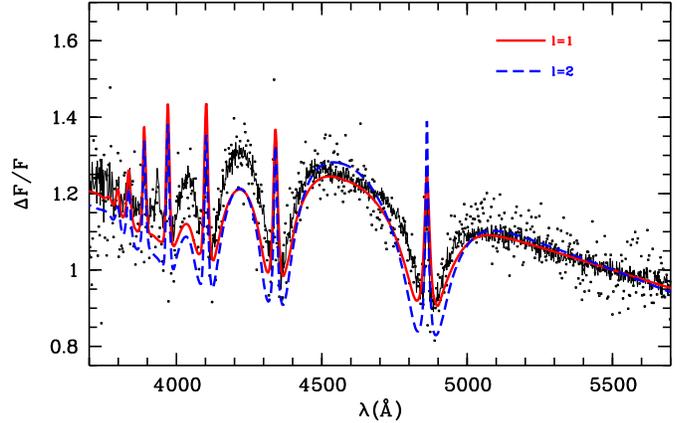}
\caption{Comparison between observed and model chromatic amplitudes for
the strongest mode of HS 0507B (F2, dots) and that of ZZ Psc 
(F1, full line) shown over a 2000\,{\AA} range only for clarity. F1 of 
ZZ Psc was identified as having $\ell=1$ by \citet{cvkw:00}. The model 
chromatic amplitudes overlaid on the data are for T$_{\rm{eff}}=12000$\,kK, 
$\log g=8$ and ML2/$\alpha=0.6$. All amplitudes have been normalised such that 
they equal 1 at 5500\,{\AA}. The data for ZZ Psc are from \citet{cvkw:00}.}
\label{fig:hszzch}
\end{figure}

\cite{cvkw:00} were able to see a clear contrast between the chromatic amplitudes
for their $\ell=1$ modes and that for the one $\ell=2$ mode present in their data 
for ZZ Psc. We see no such clear contrasts for HS 0507B 
(Fig. \ref{fig:hschamp}). Mostly, this simply reflects the fact that HS 0507B is 
about 8 times fainter, and that our signal-to-noise ratios are 
correspondingly lower. 
In Fig. \ref{fig:hszzch}, we compare the observed chromatic amplitudes 
for the strongest modes in HS 0507B (F2) and ZZ Psc (F1, $\ell=1$). We overlay model 
chromatic amplitudes for comparison. These were computed using model atmospheres 
\citep[kindly provided by D. Koester; earlier versions described in][]{fin:97}. 
The similarity of the displayed chromatic amplitude of HS 0507B and ZZ Psc and the 
dissimilarity of either to the models is striking. Pending improvements to 
the models, we must invent other methods to distinguish between $\ell=1$ and $\ell=2$ 
modes.

A careful inspection of Fig. \ref{fig:hschamp} reveals that F2, F3, F4, and F6 
look qualitatively similar to each other while the continuum shape and curvature 
-- especially between H$\beta$ and H$\gamma$ -- are somewhat different for F7. 
Although the continuum between the line cores is not smooth for the weaker
multiplets, the general shape of the line cores is not unlike that of the 
stronger modes. We focus upon these differences -- the same ones found for
ZZ Psc -- in order to glean any clues on the spherical degree of the 
pulsation modes. 

For ZZ Psc, two distinguishing features of the $\ell=1$ modes compared to the 
one $\ell=2$ mode were that the latter had a larger curvature in the continuum
regions between the line cores and that the overall slope was steeper.
In the hope of separating out modes having different $\ell$ values, we attempt to 
measure these two quantities by fitting a 2$^{\mathrm{nd}}$-order polynomial 
between the lines cores as differences in these regions are readily apparent. We 
use ZZ Psc as a test case as at least one $\ell=2$ mode was identified by simple 
inspection.

In order to minimize the effect of local amplitude variations on the
curvature and to minimize the covariance between the various
parameters, we fit $y = y_0+y_{1}z+y_{2}z^{2}$, where $y=\ln a$ is the
(natural) logarithm of the amplitude $a$ and
$z=\ln\lambda-\langle\ln\lambda\rangle$, with
$\langle\ln\lambda\rangle$ the average of $\ln\lambda$ in the
wavelength region of interest.  Here, we only show the results for the
region between H$\beta$ and H$\gamma$ (4370--4820\,{\AA}). We also define 
$y_{0m}= (y_{0}-\langle\ln a\rangle_{5500})/
(\langle\ln\lambda\rangle-\ln5500)$. In this scheme, $y_{0m}$ can be
seen to be a measure of the slope of the entire spectrum, while
$y_{1}$ measures the local slope. Plotting $y_{2}$ against $y_{0m}$
should group together the modes that have similar shapes and
curvatures.

\begin{figure}[!t]
\plotone{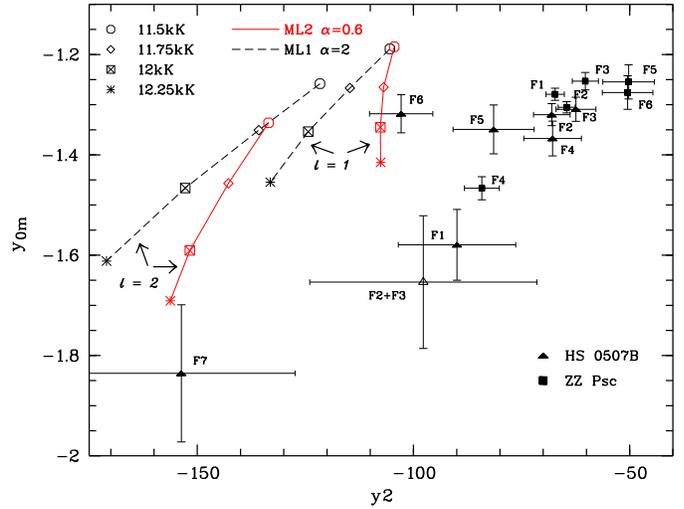}
\caption{Measures of the observed (HS 0507, ZZ Psc) and
model curvatures in the chromatic amplitudes between H$\beta$
and H$\gamma$. Energy transport by convection is comparable in
the ML1/$\alpha$=2 and ML2/$\alpha$=0.6 versions. The model amplitudes 
were convolved with a Gaussian having a FWHM of 4.6\,{\AA} to match 
the seeing profile and extended over the 8$\farcs$7 slit.}
\label{fig:mixmod}
\end{figure}

Indeed Fig. \ref{fig:mixmod} shows that F4 from the \citet{cvkw:00} data set
clearly stands out from the cluster of $\ell=1$ modes. The situation is somewhat 
less clear-cut for HS 0507B: while F2, F3, F4, and F5 are consistent with 
each other and with F1 and F2 of ZZ Psc, the position of F6 
is puzzling, although its value of $y_{0m}$ is consistent with that of the
other strong modes. 
This, together with the general appearance of the chromatic amplitudes 
implies that F2, F3, F4, and F5 have $\ell=1$ and as F1 and F6 are members of
known $\ell=1$ multiplets, they too must have $\ell=1$, even though this is not 
obvious from their respective locations in Fig. \ref{fig:mixmod}.
The difference in appearance of F7 mentioned above is borne out by the 
measured slope and curvature. Given its low amplitude, we cannot unfortunately
rule out the contribution of random noise. 
We note with interest that Handler (private communication) finds a 
peak of $\sim\,0.19$\% at 3489.09\,$\mu$Hz i.e. at approximately the same 
splitting as that observed for the 355\,s and 445\,s multiplets suggesting that 
F7 may be part of an $\ell=1$ triplet. 

We can check the reliability of the above method by repeating the procedure for 
those combination modes for which the $\ell$ character of the combination is known.
The spherical harmonic character of combination modes can be deduced from selection
rules whenever the $\ell$ and $m$ values of the constituent real modes is known.
Thus for combinations involving two $m=1$ or $m=-1$ modes (for $\ell=1$), the resulting 
wavelength-dependent pulsation amplitude should have a surface distribution described by
a $Y_{2}^{\pm2}$ component and its position in Fig. \ref{fig:mixmod} should coincide
with F4 of ZZ Psc (which has $\ell=2$). Unfortunately, we do not have any independent 
knowledge of $m$ for any of the modes in ZZ Psc and cannot use its combination
modes to verify the locations of the real modes in Fig. \ref{fig:mixmod}.
Of our combination modes, F2+F3 is the largest amplitude one that satisfies the 
above $m$-requirement. Its location in Fig. \ref{fig:mixmod} is roughly consistent 
with it having an $\ell=2$ component, although the uncertainties are too large
for the measurement to be significant. In general, our conclusion from the 
above exercise is that although one obtains quantitative measures for the differences
between modes having differing $\ell$ indices, it merely serves to corroborate
the conclusions arrived at by visual inspection of the chromatic amplitudes.

As an aside, we note that using all the real modes listed in Table A1 in \citet{vkcw:00}
results in one other mode, F8 ($920\pm3$\,s), being almost coincident with the
position of the $\ell=2$ mode (F4, 776\,s) in Fig. \ref{fig:mixmod}, while FA (500\,s)
is consistent with the position of F4 within the errors. Taken at face value, this implies
that the 920\,s mode also has $\ell=2$. \citet{kman:98} assumed $\ell=1$ for their 918\,s 
mode so it would be interesting to check the validity of this assumption by assigning
a value of $\ell=2$ to this mode in the pulsation models and attempting to quantify 
the resulting changes, if any, to the derived properties of ZZ Psc. Given the 
relatively low amplitude of F8 (0.47\%), other independent constraints would, of course,
be desirable.

We repeated the above procedure on the model chromatic amplitudes for a range of 
temperatures and a number of different sets of mixing-length parameters. We show 
two examples in Fig. \ref{fig:mixmod} one for ML2/$\alpha=0.6$, which was found 
to yield the best description of the average optical-to-ultraviolet spectra 
\citep{berg:95} and one for ML1/$\alpha=2.0$, which lies in the parameter space
of models with which the observed magnitude difference between HS 0507A and B was 
best reproduced \citep{jord:98}; see \citet{jord:98} for a description of the 
terminology. We find that all models, like the two shown, have systematically higher 
curvature than the observations. Thus, we extend to all models the conclusion of 
\citet{cvkw:00} that while the salient features of the chromatic amplitudes are 
reproduced by the models, the details leave much to be desired. It may well be that 
better agreement will only be reached once a better description of convection becomes
available.

\section{Comparison with Pulsation Theory}
\label{sec:confront}

In this section, we attempt to place our observations within a
theoretical framework by comparing various quantities derived from
our observations to those derived from theoretical considerations by
\citet{wg:99}, \citet{gw:99a}, and \citet{incl:00}. Our primary aim in 
this section is to check whether theoretical expectations are borne out 
by the observations for both real and combination modes.
Our secondary aim is to use our observations to perform consistency
checks on some of the theoretical parameters used in the above studies.
Specifically, these parameters are the thermal time constant of the
convection zone ($\tau_{c_0}$), and the parameter ($|2\beta + \gamma|$) 
that depends on the depth of the convection zone. We additionally check
whether the inclination angle derived by \citet{gh:00} is consistent with 
our data. Before proceeding, we stress that any comparison with the theory 
is rendered difficult by a number of factors, in particular, by unresolved 
multiplets and by the presence of real and combination modes of low 
amplitude.

\subsection{Real Modes}

Following \citet{vkcw:00}, we measure the relative flux and velocity
amplitudes and phases of the real modes with $R_{V} =
A_{V}/(2\pi fA_{L})$ and $\Delta\Phi_{V} = \Phi_{V} - \Phi_{L}$, respectively.
In Table \ref{hstab}, we list the values of $R_{V}$ for all real modes. 
Here, one should bear in mind that the detection of even the strongest mode is 
only marginal. For the weaker modes, the $R_V$ values represent upper limits.

%

As a ZZ Ceti type white dwarf cools, the depth of the convection zone increases 
and longer period modes are excited. However, the flux perturbations are also more 
effectively attenuated so the emergent flux variations at the photosphere for a mode 
of fixed internal pulsation amplitude are reduced. Now turbulent viscosity in the 
convection zone ensures negligible vertical velocity gradients, making the horizontal 
velocities effectively independent of depth within the convection zone 
\citep{brick:90,gw:99b}. Thus at a fixed frequency, $R_V$ should be smaller in white 
dwarfs with thinner convection zones, i.e. the hotter pulsators. Similarly, for 
different modes within the same object, the flux attenuation increases with increasing 
mode frequency, while velocity variations pass undiminished through the convection zone. 
Thus $R_{V}$ is expected to increase with increasing mode frequency. $\Delta\Phi_{V}$ 
is expected to be equal to 90$\arcdeg$ for the adiabatic case (flux leads maximum 
positive velocity by $\pi/2$) and to tend towards 0$\arcdeg$ with increasing frequency,
as the flux variations are increasingly delayed.

\begin{figure}[!t]
\plotone{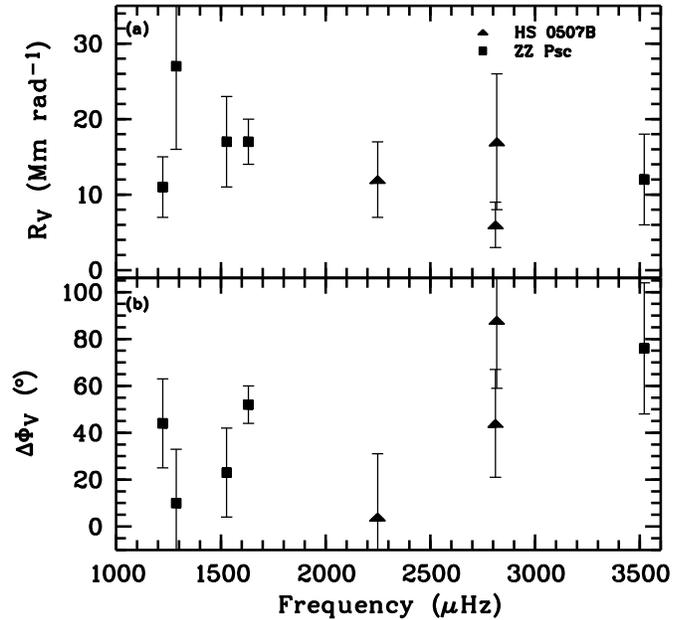}
\caption{The relative velocity to flux amplitudes \textbf{(a)} and phases 
\textbf{(b)} for HS 0507B and ZZ Psc. The values for ZZ Psc are 
taken from Table 1 in \citet{vkcw:00}.}
\label{fig:hszzrvpv}
\end{figure}

In Fig. \ref{fig:hszzrvpv}, we show the observed values of $R_{V}$ and 
$\Delta\Phi_{V}$ for both HS 0507B and ZZ Psc. The observations are broadly
consistent with expectations, in that the average values of $R_{V}$ are
reasonable, and values of $\Delta\Phi_{V}$ are between 0 and 90$\arcdeg$.
We find no clear evidence, however, for the expected trends with frequency
as the uncertainties are too large.

The longest period observed mode can be used to deduce a rough lower limit to
the value of $\tau_{c_0}$, the thermal time constant of the convection
zone, as the overstability criterion $\omega\tau_{c_0} \ga 1$ must be
satisfied for driving to exceed damping \citep{gw:99a}. Using F4 from Table 
\ref{hstab} yields a value for $\tau_{c_0}$ of $\simgt$118\,s.

\subsection{Combination Frequencies}
\label{sec:combfreq}

In models, the sharp maxima and shallower minima that typify the light curve are 
the result of the variation in the depth of the convection zone in response to the 
perturbations; this variation distorts the flux variation and gives rise to linear 
combinations of frequencies when transformed into frequency space 
\citep{brick:92,incl:00}. Thus, combination modes are not expected to have 
associated physical motion.

The relative photometric amplitudes and phases of the combination modes with respect 
to their constituent real modes can be expressed as 
$R_{C} = A_{L}^{i\pm j}/(n_{ij}A_{L}^{i}A_{L}^{j})$ where $n_{ij}$ is the factorial
of the number of {\em different\/} modes in the combination for 2-mode combinations
and $\Delta\Phi_{C} = 
\Phi_{L}^{i\pm j} - (\Phi_{L}^{i} \pm \Phi_{L}^{j}$) respectively. Relations linking
theoretical quantities to observed ones are provided by Eq. (15) and (20) in 
\citet{incl:00}. We quote these here for convenience:
\begin{equation}
\Delta\Phi_C = \tan^{-1}\left({1\over{(\omega_i \pm \omega_j) \tau_{c_0}}}\right)
\label{eq:phic}
\end{equation}
\begin{equation}
R_C = { {|2 \beta + \gamma| (\omega_i \pm \omega_j) \tau_{c_0}}
\over{ 4 \alpha_V \sqrt{1+ ((\omega_i \pm \omega_j) \tau_{c_0})^2}}}
{{ G_{\ell_i \,\, \ell_j}^{m_i\pm m_j}}\over{g_{\ell_i}^{m_i} g_{\ell_j}^{m_j}}} 
\label{eq:trc} 
\end{equation}
Here, $\omega$ is the angular frequency of the (real) mode, $\beta$ and 
$\gamma$ are related to the entropy variation, the response of the superadiabatic 
layer to the pulsation, and to the thermal relaxation time at a given depth in the 
convection zone.
${{ G_{\ell_i \,\, \ell_j}^{m_i\pm m_j}}/{g_{\ell_i}^{m_i} g_{\ell_j}^{m_j}}}$
is the geometrical multiplication factor which depends on the inclination,
$\Theta_{0}$ between the pulsation axis and the line of
sight\footnote{Values of this factor for grey atmospheres and 
$\ell_{i}=\ell_{j}=1$ are listed in Table 3 of \citet{incl:00}. 
It is equal to $0.65+0.45/\cos^2\Theta_0$ for $m_i=m_j=0$  combinations,
$0.65+0.90/\sin^{2}\Theta_0$ for $m_i=-m_j=\pm1$ combinations, and
0.65 for all others.}
and $\alpha_V \sim 0.4$ is a factor relating photometric amplitudes
in the V band to bolometric ones.

In what follows, we take the inclination angle between the pulsation axis and
the line-of-sight to be 79$\arcdeg$ as inferred by \citet{gh:00}. We caution, 
though, that its derivation assumed not only that the pulsation axis is
closely aligned to the rotation axis of the star, but also that multiplets have 
the {\em same\/} intrinsic amplitude. While the former assumption is probably 
justified, the latter is not self-evident as the amplitude of the modes may depend 
highly non-monotonically on frequency \citep[especially if they are determined 
by parametric instability e.g.] []{wg:01}.  Observationally, if the intrinsic amplitudes 
of the multiplet components are identical, one would expect the $m=\pm1$ components 
of the triplets to have the same amplitude, and the ratio of the $m=0$ to the $m=\pm1$ 
to be the same for different multiplets. While the observations are generally 
consistent with the above, there is a glaring exception: the $m=\pm1$ components of 
the 355\,s triplet, which includes the strongest mode, have very different 
amplitudes -- both in our data (F1, F2), and in those of \citet{gh:00}.
The derived value of $\Theta_{0}$ should therefore only be considered as approximate.

\begin{figure}[t]
\plotone{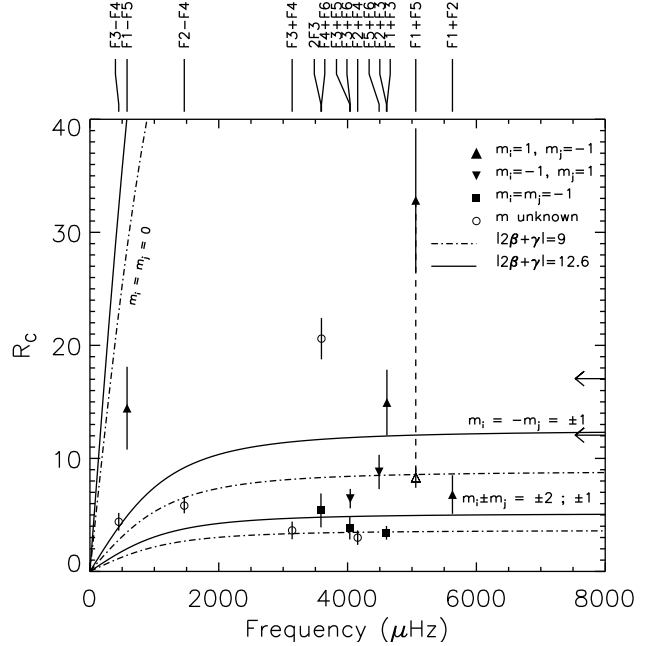}
\caption{Measure of strength of two-mode combinations,
$R_{C} = A_{L}^{i\pm j}/(n_{ij}A_{L}^{i}A_{L}^{j})$, as
a function of frequency.
Filled and open squares denote those combinations for which
values of $m$ are known \citep[from][]{gh:00} and unknown respectively.
Theoretically expected values of $R_{C}$ for $\ell_{i}=\ell_{j}=1$ are 
calculated using Eq. (\ref{eq:trc}) with an inclination angle of 79$\arcdeg$ 
and $\tau_{c_{0}} = 118$\,s. The assumed value for $|2\beta + \gamma|$ is
indicated. The upper and lower arrows denote the asymptotic value of 
$m_i=-m_j=\pm1$ combinations with $\Theta_0=50$\arcdeg\ in Eq. (\ref{eq:trc}), 
with $|2\beta + \gamma|=12.6$ and 9 respectively. It is an example of how other 
values of $\Theta_0$ would shift the theoretical curves. (As indicated
in the text, the $m_i\pm m_j = \pm2,\pm1$ curve is independent of $\Theta_0$).
The dashed line ending in the open triangle shows where the F1+F5 
combination would be shifted to if it has been misidentified (see text and 
Table \ref{hstab}).} 
\label{fig:theorrc}
\end{figure}
\begin{figure}[t]
\plotone{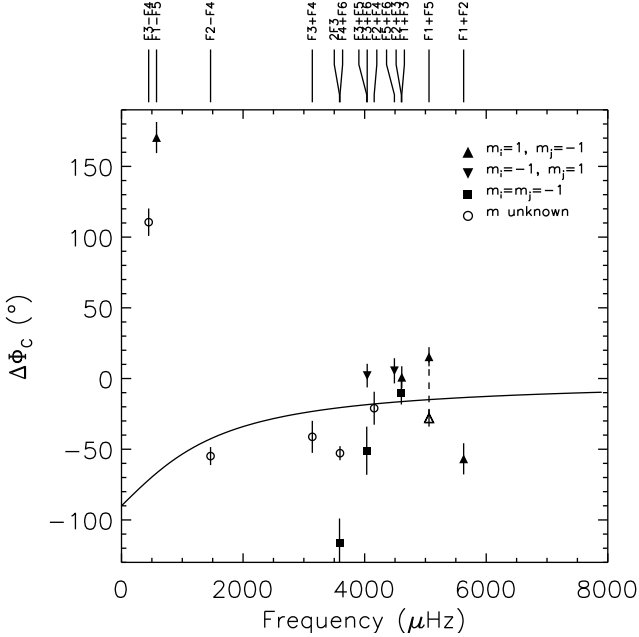}
\caption{Observed (filled squares) values of $\Delta\Phi_{C}$ and the
theoretically expected values as a function of frequency, as obtained
from Eq. (\ref{eq:phic}), assuming a value of 118s for $\tau_{c_0}$.
As in Fig. \ref{fig:theorrc}, the dashed line ending in an open
triangle, shows the position of F2+F6.}
\label{fig:theorphic}
\end{figure}

Figure \ref{fig:theorrc} shows the observed and theoretically expected
values of $R_{C}$. For the latter, we used $\Theta_{0} = 79\arcdeg$,
$\tau_{c_{0}}=118$\,s (as derived from the longest period overstable
mode) and $|2\beta+\gamma|=12.6$, as estimated from theoretical
arguments by \citet{incl:00}.
Given the numerous assumptions that have to be made in 
deriving theoretical values of $R_{C}$ and the uncertainty in the measured
mode amplitudes, especially at lower frequencies, the agreement is adequate:
values of $R_{C}$ are expected to be lower for lower frequency combinations
($\omega\tau_{c_0} \simeq 1$) and almost constant for higher frequency
combinations ($\omega\tau_{c_0} \gg 1$) as can be seen from Eq. (\ref{eq:trc}). 
The best agreement with expected values of $R_{C}$
is shown by combinations involving two $m = -1$ modes, these are
also the largest amplitude modes in general.

For combinations having $m_{i}\pm m_{j} = \pm2;\pm1$, the theoretical
value of $R_{C}$ is independent of the inclination of the pulsation
axis to the line-of-sight ($\Theta_0$) and depends only on the assumed 
value of $|2\beta + \gamma|$, while for $m_{i}=-m_{j}=\pm1$ combinations,
$R_{C}$ is dependent on $|2\beta + \gamma|$ and $\Theta_0$, but is
independent of $\tau_{c_0}$ for higher frequency combinations ($\omega
\tau_{c_{0}} \gg 1$). Taken together, $\beta$ and  $\gamma$ represent a 
measure of the depth of the convection zone as a function of effective 
temperature of the white dwarf and thus relate directly to the width of 
the instability strip: the lower $|2\beta+\gamma|$ is, the wider the 
instability strip \citep{incl:00}. Figure \ref{fig:theorrc}
shows that better agreement with the observed values of $R_{C}$ is 
obtained using a somewhat lower value of 9 for $|2\beta + \gamma|$.
For this value, the $m_i=-m_j=\pm1$ modes are also better reproduced.
This supports a high value for the inclination angle: for smaller
$\Theta_0$, the theoretically expected $R_{C}$ values for
$m_{i}=-m_{j}=\pm1$ would shift upwards and away from the observations.

The relative phases are shown in Fig. \ref{fig:theorphic}. These
are expected to be in phase with their corresponding real modes i.e.
$\Delta\Phi_{C}\sim0$. While many combinations follow the expected
trend, there are some large deviations; we discuss this further below. 

In the above, we used $\tau_{c_0}=118\,$s, as derived from the 
longest period overstable mode. In principle, one might improve on this by 
fitting the whole set of combination modes for $\tau_{c_0}$, $|2\beta+\gamma|$
and $\Theta_0$ simultaneously. However, \citet{incl:00} showed that
for pairs of sum and difference combination frequencies arising from
the {\em same} real modes (e.g. F1$\pm$F5), the ratio of the amplitudes
does not depend on $|2\beta+\gamma|$ and thus one could derive
$\tau_{c_0}$ directly.  We quote Eq. (21) from the above-cited paper:
\begin{equation}
 {A_L^{i+j}\over A_L^{i-j}}
= {{(\omega_i + \omega_j)}\over{(\omega_i -
\omega_j)}} {{\sqrt{1+ (\omega_i - \omega_j)^2 \tau_{c_0}^2}}\over
{\sqrt{1+ (\omega_i + \omega_j)^2 \tau_{c_0}^2}}}
{{ G_{\ell_i \,\, \ell_j}^{m_i+m_j}}\over{G_{\ell_i \,\, \ell_j}^{m_i-m_j}}}
\label{eq:combrat}
\end{equation}
Note that the geometrical factor and hence the dependence on $\Theta_0$ 
cancels out if at least one constituent real mode has $m=0$.

From the list of combination frequencies in Table \ref{hstab}, there
are three pairs that may be used, viz., F1$\pm$F5, F3$\pm$F4, and
F2$\pm$F4. None of these have an $m=0$ component, but we can still
solve Eq. (\ref{eq:trc}) for $\tau_{c_0}$ using the estimate of
$\Theta_0=79\arcdeg$ of \citet{gh:00}. The results are listed in
Table \ref{tab:tauc}; we see that the values of $\tau_{c_0}$ thus
derived are compatible with the value ($\sim\!118$\,s) derived from
the longest period real mode excited. Using $\Theta_0=79\arcdeg$
together with Eq. (\ref{eq:trc}) and \ref{eq:combrat} yields
a value of $\tau_{c_0}$ that is entirely consistent with the value
found from the longest period real mode. This agreement lends some
confidence to Eqs. (\ref{eq:trc}) and (\ref{eq:combrat}). 
We therefore point out that only the choice of $m=-1$ for F4
resulted in physically acceptable solutions; this is consistent with 
all observed multiplet components in our data having $m\ne0$ and with 
the indication from the chromatic amplitudes (Sect. \ref{sec:champ}) that 
all modes that give rise to combinations have $\ell=1$. As we point out 
below, the F1+F5 combination may in fact be F2+F6. In view of this, it is 
hardly surprising that the F1$\pm$F5 pair came to naught.
\begin{table}[h]
\caption[]{Values of $\tau_{c_0}$ derived from combination pairs}
\label{tab:tauc}
\begin{centering}
\begin{tabular}{cll}
\hline
      Combination pair &   & $\tau_{c_0}$ (s) \\
\hline
       F1$\pm$F5        &   &  \nodata                   \\
       F2$\pm$F4        &   &  112 ($m = -1$)              \\
       F3$\pm$F4        &   &  122 ($m = -1$)             \\
\hline
\end{tabular}
\end{centering}
\tablecomments{For the F2,3$\pm$F4 pairs, $\ell_{j}=1, m=1,0$ resulted in
unphysical solutions. Likewise for the F1$\pm$F5 pair. All possible
permutations of $\ell_{i}=1, m_{i}=-1, \ell_{j}=2, m_{j}=\pm 2, \pm 1, 0$
gave unphysical solutions for the F2$\pm$F4 pair. Likewise for F3$\pm$F4 except
for $\ell_{j}=2, m_{j}=-1$, for which $\tau_{c_0}$=19\,s.}
\end{table}

In Figs. \ref{fig:theorrc} and \ref{fig:theorphic}, there are a number
of combinations for which $R_{C}$ and $\Delta\Phi_{C}$ are rather
discrepant from the theoretically expected values. These
discrepancies might simply be due to unresolved, low amplitude modes,
to which the measured phases in particular are rather sensitive.
However, some of the anomalies can be explained more easily, as
(i) resulting from degenerate combinations\footnote{Degenerate
combinations are a particular problem for HS 0507B, since the frequencies
of several real modes have near-integer ratios e.g. F3:F4 $\sim$ 4:3 
\citep[see Table \ref{hstab} and][]{gh:00}.} that are an inevitable
consequence of equal frequency splittings, and (ii) the possibility for
low amplitude real modes to have high amplitude combinations.
We discuss each of these possibilities below, using our observed
combination modes as examples.

A case in point is our F1$+$F5 combination that lies very close
in frequency to the F2$+$F6 combination. In hindsight, the anomalous
value of $R_{C}$ points to a misidentification, but only because we
now have an $\ell$ value for the F1 and F5 modes, bolstered by an
estimate of the inclination angle. Including F2+F6 in the fit of the Fourier 
Transform of the light curve (Sect. \ref{sec:periodicities}) instead
of F1+F5 results in a value of $R_{C}$ (8 $\pm$ 1) that is more compatible 
not only with the theoretically expected value, but one that is also more in
line with other combinations involving $m_i=1, m_j=-1$ (see Fig.
\ref{fig:theorrc}). The value of $\Delta\Phi_{C}$, $-28 \pm 5\arcdeg$, is also
in better agreement (see Fig. \ref{fig:theorphic}).\footnote{Our justification
for initially preferring F1+F5 is as stated in Sect. \ref{sec:periodicities} 
i.e. that we obtained a slightly larger fitted amplitude and a 
correspondingly lower $\chi_{\mathrm{red}}^{2}$ for F1+F5.}

More problematic are combinations that are degenerate with
$m_i=m_j=0$ combinations, as with $\Theta_0=79\arcdeg$ these
combinations have very high $R_C$ values (see Fig.~\ref{fig:theorrc}),
as a result of strong cancellation for the real mode and little
cancellation for the combination (which has an $\ell=0$ component).
This means that even real modes with weak observed amplitudes can
give rise to relative strong combinations\footnote{Indeed, for
$\Theta_0=90\arcdeg$, the combination mode would be visible even
though the corresponding real modes would not! Obviously, this could
lead to gross misidentifications.}. As an example, consider our F1+F3
($m_i=1, m_j=-1$). This combination has the same frequency as f3+f9
i.e. $1796.70+2814.02\,\mu\mathrm{Hz}$, $m_{i}=m_{j}=0$ -- of
\citet{gh:00}. Fitting our Fourier Transform using the frequencies of
\citet{gh:00} (Sect. \ref{sec:periodicities}), we find that the sums of the
phases of the two combinations of real modes are almost the same:
$-29\arcdeg$ for f3+f9 and $-32\arcdeg$ for F1+F3. Hence, these
combinations will add coherently. If the real modes in the
different multiplets have the same {\em intrinsic} amplitude, one
would expect both combinations to have roughly the same {\em observed}
amplitude, and hence this could plausibly lead to an anomalously
high value of $R_C$ for F1+F3. 

Another such combination is F$4_{m=-1}$+F$6_{m=1}$ (1346.1+2248.4\,$\mu$Hz)
identified by Handler (private communication) as $f_{2_{m=-1}} + f_{4_{m=1}}$ 
(1793.29+1800.7\,$\mu$Hz).
It has an amplitude that is not only the largest among the combination 
modes but is also larger than that of 2 of our real modes (F1 and F7).
Another combination that also adds up to $\sim$3594\,$\mu$Hz is the
first harmonic of the $m=-1$ component of the 556\,s multiplet i.e.
$2\times 1796.7$ or $2f_3$ in \citet{gh:00}. As mentioned earlier,
even though the $m=0$ components are weak, their harmonics are not.
This, coupled with the integer frequency ratios for the modes involved
means that different combinations can (and do) have degenerate frequencies.
In this respect, it may be interesting that the one mode from the study of 
\citet{gh:00} that changed in amplitude is $f_{4_{m=1}}$ (Fig. \ref{multipdft}):
perhaps this is the result of a resonance between these modes. This could
explain the anomalously high value of $R_{C}$ for the F4+F6 combination.

In summary, therefore, we find that most of the apparently discrepant 
values of $R_C$ and $\Delta\Phi_C$ can be explained by the contribution of
combinations of unresolved or low-amplitude modes, degenerate combinations,
and low frequency noise.
\section{Squaring the circle}

We have used high signal-to-noise, time-resolved spectra in an attempt 
to identify the spherical degree of the pulsation modes and 
to infer some of the properties of the outer layers of a pulsating white 
dwarf. In the process of doing so, we have tested several aspects of 
the very theory that was used in the analysis. Real and combination modes 
were used in conjunction with wavelength-dependent fractional amplitudes 
and velocity amplitudes in order to piece together a consistent picture. 
At almost every step, our results were compared with those of ZZ Psc, a 
brighter and better-studied white dwarf.

Our measured frequencies and amplitudes (see Table \ref{hstab}) for 
all modes except one are in agreement with those reported by \citet{gh:00}.

We measured the line-of-sight velocities associated with
the pulsations and found a marginally significant modulation at the frequency 
of the strongest mode. This can be taken as a detection of surface motion in 
a second ZZ Ceti type pulsator, or, more conservatively, as a stringent upper 
limit to such motion. From all modes, we find that the average value of (or 
upper limit to) the ratio of the velocity to flux amplitudes, $R_V$, is just
over half of that observed in ZZ Psc. The difference in the thermal time 
constant of the convection zone, $\tau_{c_0}$, for the two white dwarfs of about 
a factor of two \citep[118\,s c.f. 250\,s,][]{incl:00} translates into a factor
of two in $R_V$ as the latter roughly scales with $\tau_{c_0}$ \citep{gw:99a,gw:99b}.
Our measured values of $R_V$ are therefore consistent with expectations.

Based on the general appearance and shape of the continuum between
the line cores of the chromatic amplitudes, we deduced that modes F1 - 
F6 were consistent with having $\ell=1$. This adds some confidence to the 
pulsation models of \citet{kman:98} that assign $\ell=1$ identifications to 
modes having periods of 284\,s, 355\,s, 445\,s, and 555\,s. 

The chromatic amplitudes of these $\ell=1$ modes were found to be more
similar to the $\ell=1$ modes in ZZ Psc than to those expected from
the models, supporting the conclusions of \citet{cvkw:00}. In
order to make our results more quantitative, we devised a measure of
the curvature and slope of the chromatic amplitudes and used this in
an attempt to separate $\ell=1$ and $\ell=2$ modes for both the observations
and the models (Fig. \ref{fig:mixmod}). Although this procedure did not
alter our mode identifcation based on simple inspection for HS 0507B, it
did serve to quantify differences between the models and observations.
Furthermore, we found that it worked very well for ZZ Psc and that it
yielded two additional potential $\ell=2$ modes, at 920 and 500\,s.

The relative amplitudes of the combination modes proved to be a useful tool,
as these in general follow expected trends for given $\ell$ and $m$.
Exploiting this, we attempted to redetermine $\tau_{c_0}$ using combination
frequency pairs, and obtained good agreement with the value derived 
using the longest period real mode.
A by-product of this exercise was an indication that F4 has $m=-1$ as no 
other $m$ value -- including those combinations arising from $\ell_{i}=1, 
\ell_{j}=2$ combinations -- yielded physically acceptable values of $\tau_{c_0}$,
thereby supporting our $\ell$ identification for F4. Thus, combination frequencies 
can potentially provide indirect constraints on $\ell$ and $m$ values even when 
no splittings are observed. Taking $\tau_{c_0}$ at face-value, we
infer that HS 0507B has a shallower convection zone than ZZ Psc, consistent 
with HS 0507B being slightly hotter.

The phases of the real and combination modes were slightly more difficult to 
reconcile with theoretical expectations. For combination modes this is likely 
to be due to unresolved modes that plague a relatively short 
time series. Data from longer, uninterrupted time series, for 
instance, from the Whole Earth Telescope is probably better suited to such 
analysis. We showed that weak, unresolved modes are probably being 
manifested in the amplitudes and phases of the combination modes. For the 
real modes, although our measured values of $\Delta\Phi_{V}$ are less than 
90$\arcdeg$, the trend with frequency is opposite to that expected; this 
is also true for ZZ Psc \citep{vkcw:00}.

Potential future work could involve focusing on using the observed 
chromatic amplitudes to calibrate the temperature stratification in 
current white dwarf atmosphere models and may provide insight into 
areas where these models fail to reproduce the observations, while 
attempts to match the stringent constraints of the observed period 
structure and $\ell$ and $m$ identifications using pulsation models 
may yield a unique asteroseismological solution for HS 0507B.

\begin{acknowledgements}
We are indebted to G. Handler for making his results available
to us prior to publication and to Y. Wu for many clarifications of
the theoretical aspects of this study.
We also acknowledge support for a fellowship of the Royal Netherlands
Academy of Arts and Sciences (MHvK) and partial support from the
Kungliga Fysiografiska S\"{a}llskapet (RK). R.K. would additionally
like to thank J. S. Vink for encouragment, and 
the Sterrenkundig Instituut, Utrecht for its hospitality.
This research has made use of the SIMBAD database, operated at CDS,
Strasbourg, France.
\end{acknowledgements}

\appendix
\section{Corrections for instrumental flexure and differential refraction}
\label{sec:flexref}
\begin{figure}[t]
\plotone{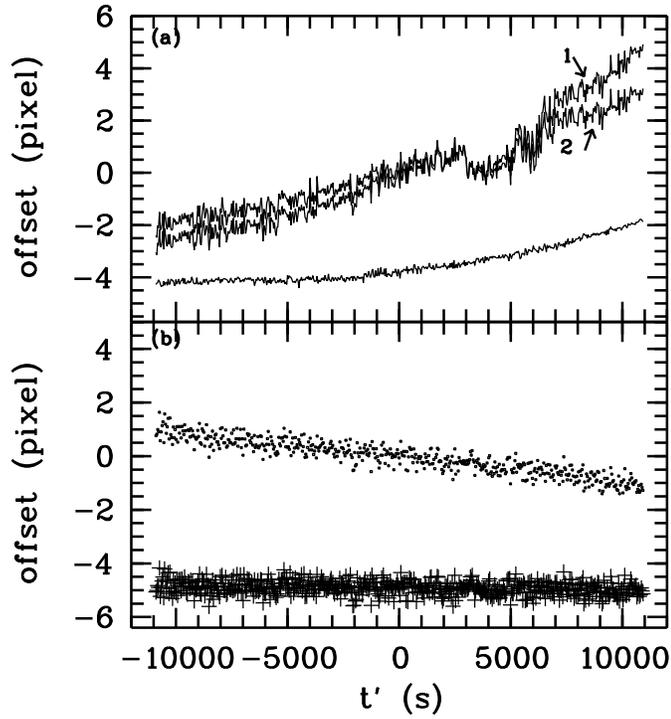}
\caption{\textbf{(a)} The lowest curve (offset by 3.5 pixels) shows the pixel positions 
         of the \ion{O}{i} $\lambda$ 5577\,{\AA} line as a function of time. The upper 
         curve (1) shows the variations in the position of H$\beta$ while (2) shows
         the variations in H$\beta$ after correcting for instrumental flexure.
         \textbf{(b)} The dots show the resulting positions of H$\beta$ after
         shifting with respect to H$\gamma$ and the pluses the
         final positions of H$\beta$ after correcting for both flexure and refraction
         (offset by 5 pixels).
         (a) and (b) together show the various stages in the processing that lead
         to the final positions of the lines. Note that although the above figure
         is for the first night only, it is representative of the procedure for both
         nights.}
\label{fig:newoffsets}
\end{figure}

We were especially fastidious in correcting the wavelength scale
for the effects of flexure and differential atmospheric refraction.
We measured the instrumental flexure from the positions of the \ion{O}{i} 
$\lambda$5577\,{\AA} sky line which were derived by cross-correlating 
the flux-calibrated spectra using an average of all spectra as a template.
We found the positions to be adequately represented by a third-order
polynomial fit (Fig. \ref{fig:newoffsets}a).
The effect of differential refraction on line and continuum intensities is
non-negligible even at modest airmasses and increases rapidly with decreasing
wavelength. Having corrected for instrumental flexure,
the resulting positions of the Balmer lines vary linearly with
$\sin\alpha\tan\zeta$ where $\alpha$ is the difference between the parallactic
and position angles and $\zeta$ is the zenith distance. This is indicative of
the remaining shifts being dominated by refraction. (Upper curve Fig. 
\ref{fig:newoffsets}b).
We computed a value for the refractive index relative to that at H$\gamma$
using the standard prescription of \cite{stone:96} and use this value to correct
all pixels in the spectrum.
As the position of the stars in the slit was altered after
a correction for a guider malfunction (on the second night) compared to that
at the beginning of the observing run, we applied a 5.1\,{\AA} shift to the spectra
of both components. This shift was computed by measuring the wavelength
of H$\gamma$ of HS 0507A in a narrow-slit (0$\farcs$7) frame and insisting
that the average wavelength of H$\gamma$ -- as obtained from line-profile fitting --
be the same. For complete consistency, we applied a 1.74\,{\AA} shift to the spectra
of the previous night.
\section{HS 0507A}
\label{sec:acheck}
\begin{figure}[[htb]
\plotone{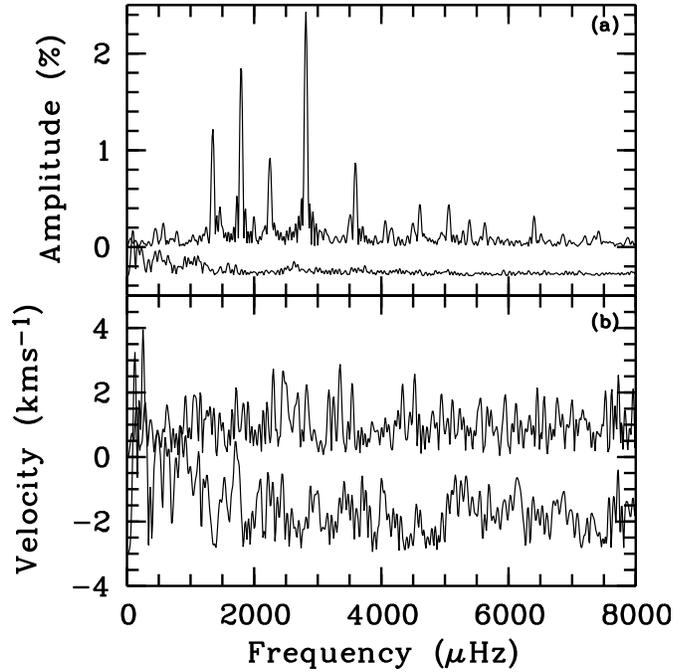}
\caption{\textbf{(a)} Fourier transforms of the light curves of HS 0507B (top) and 
HS 0507A (offset by $-$0.3\%) \textbf{(b)} Fourier transforms of the velocity curves
of HS 0507B (top) and HS 0507A (offset by $-3$\,kms$^{-1}$). The average velocity 
curve of HS 0507A, as used here, was constructed from the flexure and 
refraction-corrected positions of the three strongest Balmer lines. Polynomial fits 
were used to describe wandering in the slit for HS 0507A whereas H$\gamma$ of HS 0507A 
was used to correct for wander of HS 0507B as described in Sect. \ref{sec:obs}. The error 
in a single measurement of the velocities for HS 0507A is $\sim$ 9\,kms$^{-1}$. Note 
the lack of significant velocity signal in HS 0507A at the frequency of the strongest 
mode as obtained from the flux variations of HS 0507B. The Fourier transforms shown 
here are for the first night only.} 
\label{fig:acheck1}
\end{figure}

As was mentioned in Sect. \ref{sec:obs}, the non-variability of HS 0507A is 
crucial to the velocity measurements and the continuum variations of HS 0507B as 
these are carried out with respect to HS 0507A. The measurements of HS 0507A 
are, of course, not carried out relative to any other star, this is further elaborated 
in Fig. \ref{fig:acheck1}.
%
%
\end{document}